\documentclass{PoS}

\title{Sterile neutrinos in the 3+s scenario and solar data}

\ShortTitle{Sterile $\nu$'s and solar data}

\author{\speaker{Jo\~{a}o Pulido}\\
        Centro de F\'{\i}sica Te\'{o}rica de Part\'{\i}culas, Instituto Superior T\'{e}cnico, 
        Av. Rovisco Pais, 1049-001, Lisboa, Portugal\\
        E-mail: \email{pulido@cftp.ist.utl.pt}}

\author{C.R. Das\\
        Centro de F\'{\i}sica Te\'{o}rica de Part\'{\i}culas, Instituto Superior T\'{e}cnico,
        Av. Rovisco Pais, 1049-001, Lisboa, Portugal\\
        and\\
        Department of Physics, University of Jyv\"askyl\"{a}, Survontie 9, Jyv\"askyl\"{a}, Finland\\
        E-mail: \email{crdas@cftp.ist.utl.pt}}

\abstract{The flatness of the SuperKamiokande neutrino electron scattering spectrum and the
apparent downturn of the charged current spectrum in the SNO data which the Large Mixing Angle
solution (LMA) to the solar neutrino problem fails to predict are analysed in the context of an
extension to the standard electroweak model with light sterile neutrinos. It is found that
a sterile neutrino which is quasi degenerate with the active ones with $\Delta m^2_{41}=10^{-5}eV^2$
and mixing $sin\theta_{14}=0.04$ provides a suitable improvement to the LMA data fits.}

\FullConference{The European Physical Society Conference on High Energy Physics -EPS-HEP2013\\
		18-24 July 2013\\
		Stockholm, Sweden}

\begin{document}


The introduction of light sterile neutrinos into the Standard Model of electroweak interactions
was motivated by the experimental observations from the LSND accelerator experiment to which the
later data from other accelerator experiments like KARMEN, ICARUS, MiniBoone, along with Gallium 
calibration and reactor experiments provided additional evidence \cite{Abazajian:2012ys}. These
are very short baseline experiments ($L\sim few\times 10m$) for whose anomalous data an oscillation 
to one or two sterile neutrino states (s=1 or 2) seems to be implied with \cite{Kopp:2013vaa}
\begin{equation}
\Delta m_{41}^2,~\Delta m_{51}^2=O(|1eV^2|).
\end{equation}
Specifically, from accelerator experiments ($\nu_e$ or $\bar\nu_e$ appearance from $\nu_{\mu}$ or 
$\bar\nu_{\mu}$) it is found that 
\begin{equation}
sin^2 2\theta_{e\mu}=(4-10)\times 10^{-3},~\Delta m^2=(4-7)\times 10^{-1} eV^2
\end{equation}
while for the reactor and Gallium anomalies ($\nu_e$ and $\bar\nu_e$ disappearance)
\begin{equation}
sin^2 2\theta_{ee}=(70-200)\times 10^{-3},~\Delta m^2=(2-3)\times 10^{-1} eV^2
\end{equation}
with the definitions $sin^2 2\theta_{e\mu}=4|u_{e4}|^2|u_{\mu 4}|^2,
\;sin^2 2\theta_{ee}=4|u_{e4}|^2(1-|u_{e4}|^2)$. Owing to such a large oscillation frequency,
these sterile neutrinos do not play any role in solar neutrino oscillations. 

It is usually argued on the other hand that the solar
neutrino problem is 'solved' which is not the case. In fact, an estimation made by the Borexino
Collaboration shows that there is a gap in the knowledge of the neutrino survival probability
in the vacuum matter transition region [$O(1-5)MeV$] \cite{Collaboration:2011nga}. More importantly,
besides the long standing problem of the flatness of the SuperKamiokande (SK) spectrum 
\cite{Abe:2010hy}, \cite{Cravens:2008aa} which the Large Mixing Angle (LMA) solution fails to 
explain, also the LMA charged current (CC) spectrum prediction from the SNO experiment 
\cite{Aharmim:2009gd} seems to proceed in the opposite direction from its LMA prediction.

Investigating a survival probability leading to an electron and CC spectra more consistent with 
the SK and SNO CC ones, we were lead to introduce light sterile neutrinos and search for 
possible ranges of $\Delta m^2_{\rm new}$ and $\theta_{\rm new}$\footnote{A similar investigation 
was performed by the authors of refs.\cite{deHolanda:2003tx},\cite{deHolanda:2010am}.}. Adequate 
probability profiles were found from oscillations to sterile neutrinos which are quasi degenerate 
with respect to the active ones $(\Delta m^2_{41}=10^{-5}eV^2)$ and with small mixing to these. 
Hence they are different from the sterile neutrinos that are suggested by accelerator, reactor and 
Gallium anomalies. Our 4$\times$4 Hamiltonian describing the solar neutrino oscillations is
in the weak basis
\begin{equation}
(H_I)_W =U\left(\begin{array}{cccc} 0 & 0  & 0  & 0
\\ 0 & \frac{\Delta m^2_{21}}{2E} & 0 & 0 \\
0 & 0 & \frac{\Delta m^2_{31}}{2E} & 0 \\ 0 & 0 & 0 & \frac{\Delta m^2_{41}}{2E} \\ \end{array}\right)
U^{\dagger}+\left(\begin{array}{cccc} V_{CC} & 0  & 0  & 0
\\ 0 & 0 & 0 & 0 \\
0 & 0 & 0 & 0 \\ 0 & 0 & 0 & -V_{NC} \\ \end{array}\right)
\end{equation}
where $U$ is the straightforward 4$\times$4 extension of the usual leptonic mixing matrix, 
$V_{CC}=G_F\sqrt{2}N_e$, $V_{NC}=-G_F/ \sqrt{2}N_n$ with $N_e,~N_n$ denoting the electron and 
neutron densities. We use the representation $U=U_{34}\tilde{U}_{24}\tilde{U}_{14}U_{23}
\tilde{U}_{13}U_{12}$. At this early stage of sterile neutrino investigation for the solar case 
we assume all sterile mixings to be equal with $sin\theta_{41}=0.04$. We thus get the model 
survival probability shown in fig.1 where also the LMA probability is displayed for comparison.
\begin{figure}[htb]
\centering
\vspace{-0.3cm}
\includegraphics[height=105mm,keepaspectratio=true,angle=-90]
{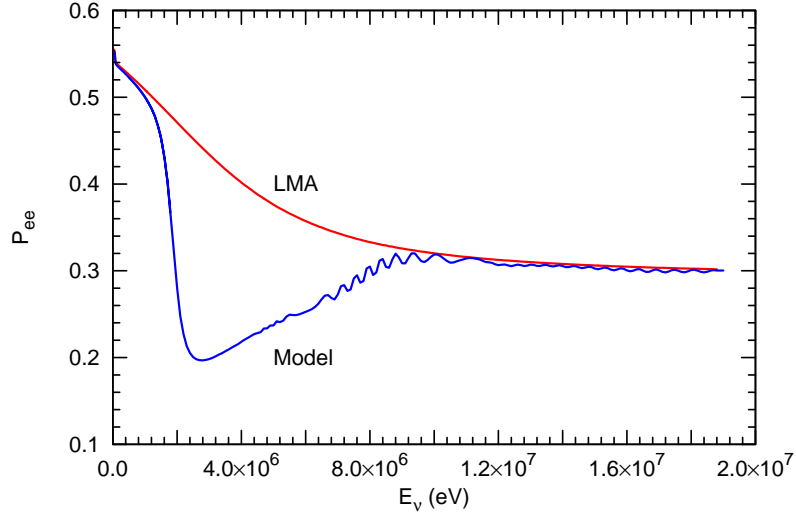}
\vspace{-0.3cm}
\caption{Electron neutrino survival probability: LMA (upper curve) and sterile model (lower
curve).}
\end{figure}
\noindent As for the relevant physical quantities, we start with the SNO CC spectrum evaluated as
\begin{equation}
R_{CC}(T_{eff})=\frac{\displaystyle\int_Q^{E_{max}}\frac{d\phi_{\nu}(E)}{dE}P(E)\int_{m_e}^{E-(Q-m_e)}R(T_{eff},T)
\frac{d\sigma_{CC}}{dT_{eff}}dTdE}{P(E)\rightarrow 1}
\end{equation} 
(see fig.2) where $Q=1.442~MeV$ and $T,T_{eff}$ are the physical and measured kinetic energy of the 
electron. In eq.(5) $R(T_{eff},T)$ is the energy resolution function and the rest of the notation is 
standard.
\begin{figure}[htb]
\centering
\vspace{-0.3cm}
\includegraphics[height=105mm,keepaspectratio=true,angle=-90]
{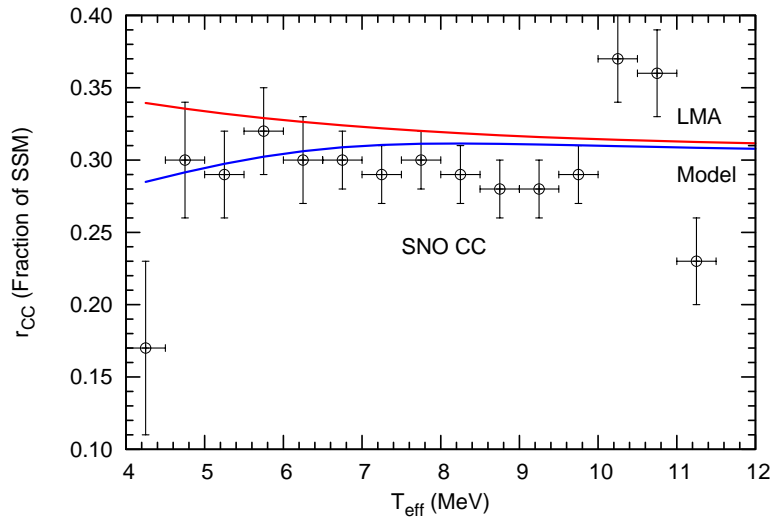}
\vspace{-0.3cm}
\caption{CC spectrum in SNO: the LMA prediction (upper curve), the model prediction (lower curve) 
and the data \cite{Aharmim:2009gd}.}
\end{figure}
For the solar neutrino fluxes we used the AGSS09ph model \cite{Serenelli:2009ww}.
We have also evaluated the electron scattering spectrum (ES) for SNO, SK and Borexino 
from the expression
\begin{equation}
R_{ES}(E_{eff})=
\frac{\displaystyle\int_{m_e}^{{E_e}_{max}}dE_e~R(E_{eff},E_e)
\int_{E_m}^{E_M}dE\phi_{\nu}(E)\left[P_{ee}(E)\frac{d\sigma_{e}}{dE_e}+
\left(P_{e\mu}(E)+P_{e\tau}(E)\right)\frac{d\sigma_{\mu,\tau}}{dE_e}\right]}
{\displaystyle\int_{m_e}^{{E_e}_{max}}dE_e~
R(E_{eff},E_e)\int_{E_m}^{E_M}dE\phi_{\nu}(E)\frac{d\sigma_{e}}{dE_e}}
\end{equation}
where $E,E_{eff}$ are the physical and measured electron energy. Its LMA and model predictions are shown 
together with the SK data from 2010 (fig.3) and 2008 (fig.4). The downturn in the CC data for the lower energies 
is clearly seen in fig.2 and a hint of the same effect can also be seen in the ES scattering data, especially
in the second set (fig.4). Such an effect which the LMA model alone fails to account for, is clearly 
predicted by the sterile model.
\begin{figure}[htb]
\centering
\vspace{-0.3cm}
\includegraphics[height=105mm,keepaspectratio=true,angle=-90]
{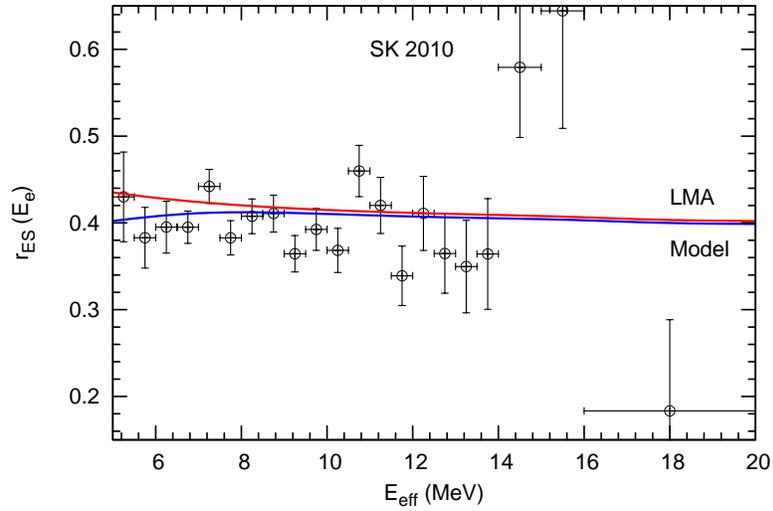}
\vspace{-0.3cm}
\caption{ES scattering in SK: the 2010 data \cite{Abe:2010hy}. The upper and lower curves are the LMA and sterile
model predictions respectively.}
\end{figure}

We also evaluated the model predictions for the total rates which are given in table I along with the LMA ones 
and the data.
\vspace{-0.1cm}
$$\begin{array}{cccccccc}
     & Ga (All, \leq  & Cl    & SNO(CC)      & SNO(NC)     & SNO(ES)      & SK           & Borexino \\ 
     & Dec.2007)      & (SNU) & (\times 10^6 & (\times 10^6 & (\times 10^6 & (\times 10^6 & (\times 10^6 \\
     & (SNU)          &       & cm^2 s^{-1})  & cm^2 s^{-1})  & cm^2 s^{-1})  & cm^2 s^{-1})  & cm^2 s^{-1}) \\ \hline 
     & 66.1       & 2.56           & 1.67                         & 5.54                         & 1.77                         & 2.32               & 2.40     \\
Data & \pm3.1     & \pm0.16        & \pm^{0.05}_{0.04}               & \pm^{0.33}_{0.31}               & \pm^{0.24}_{0.21}               & \pm{0.04}          & \pm{0.4} \\
     &            &        \pm0.15 &               \pm^{0.07}_{0.08} &               \pm^{0.36}_{0.34} &               \pm^{0.09}_{0.10} &          \pm{0.05} & \pm{0.1} \\ \hline
LMA  &  62.4 & 2.70 & 1.69 & 5.22 & 2.21 & 2.21 & 2.27 \\
Model & 61.0 & 2.60 & 1.61 & 5.13 & 2.14 & 2.14 & 2.12 \\ \hline
\end{array}$$
\begin{center}
{\bf Table I:} the data and the LMA and model predictions for the total rates. Units are in SNU for Gallium
and Chlorine and in $10^6 cm^2 s^{-1}$ for the remainder. For Ga all data from SAGE and Gallex/GNO are 
included up to Dec.2007
\end{center}
\vspace{0.4cm}
We next perform an analysis of the quality of the fits to the data. Using the standard $\chi^2$ definition 
\cite{Das:2009kw}
\begin{equation}
\chi^2=\sum_{j_{1},j_{2}}({R}^{th}_{j_{1}}-{R_{j_{1}}}^{\exp})\left[{\sigma^2}
(tot)\right]^{-1}_{j_{1}j_{2}}({R}^{th}_{j_{2}}-{R_{j_{2}}}^{\exp})
\end{equation}
where indices $j_{1},j_{2}$ run over the 7 solar neutrino experiments and the error matrix
includes the cross section, the astrophysical and the experimental uncertainties, we obtain
for the rates only, with $\Delta m^2_{sterile}$ and $\theta_{sterile}$ as free parameters,
\begin{equation}
\chi^2_{rates}(\rm{LMA})=8.1/5~d.o.f.~~~,~~~\chi^2_{rates}({\rm model})=15.3/5~d.o.f.
\end{equation}
A word of caution must be inserted here as regards the inclusion of the Ga rate, since its contribution to 
$\chi^2_{rates}$ is overwhelming. Had we taken for instance the Ga/GNO data only from the period 1998-2003 
(62.9$\pm$5.4$\pm$2.5 SNU) the result would be
\begin{equation}
\chi^2_{rates}(\rm{LMA})=5.5/4~d.o.f.~~~,~~~\chi^2_{rates}({\rm model})=12.5/4~d.o.f
\end{equation}
so $\chi^2_{rates}$ strongly depends on the Ga data period one considers. Moreover the Ga rate has
been decreasing all along its history of data taking (see table II), a fact whose origin remains unclear.

$$\begin{array}{|c|c|c|} \hline
Gallex I & \leq June~1992 & 83\pm19 \\ 
Gallex II & Aug'92\rightarrow Jun'94 & 76\pm10 \\
Gallex III & Oct'94\rightarrow Oct'95 & 54\pm 11 \\ \hline \end{array}$$

$$\begin{array}{|ccc|} \hline
  &  1991-97  &  1998-03 \\ \hline 
Gallex/GNO &  77.5\pm6.2\pm^{4.3}_{4.7} & 62.9\pm5.4\pm2.5 \\
SAGE &  79.2\pm8.6\pm^{4.3}_{4.7} & 63.9\pm5.0 \\ \hline \end{array}$$
$$\begin{array}{|cccccc|} \hline
  & 2003  & 2004 & 2005 & 2006 & 2007 \\ \hline
SAGE & 60\pm10 & 72.5\pm12.5 & 53\pm9 & 68\pm10 & 58.5\pm8.5 \\ \hline
\end{array}$$
\begin{center}
{\bf Table II:} the evolution of the Ga rate over time (units are in SNU).
\end{center}
Hence the above $\chi^2_{rates}$ values may well be meaninglessly high and misleading. Removing the
Ga rate from the calculation, one gets instead
\begin{equation}
\chi^2_{rates}(\rm{LMA})=5.7/4~d.o.f.~~~,~~~\chi^2_{rates}({\rm model})=5.9/4~d.o.f
~~~~~~[\rm{no~Ga~rate}]
\end{equation}
and thus fits of equivalent quality for LMA alone and the sterile neutrino models.
\begin{figure}[htb]
\centering
\vspace{-0.3cm}
\includegraphics[height=110mm,keepaspectratio=true,angle=-90]
{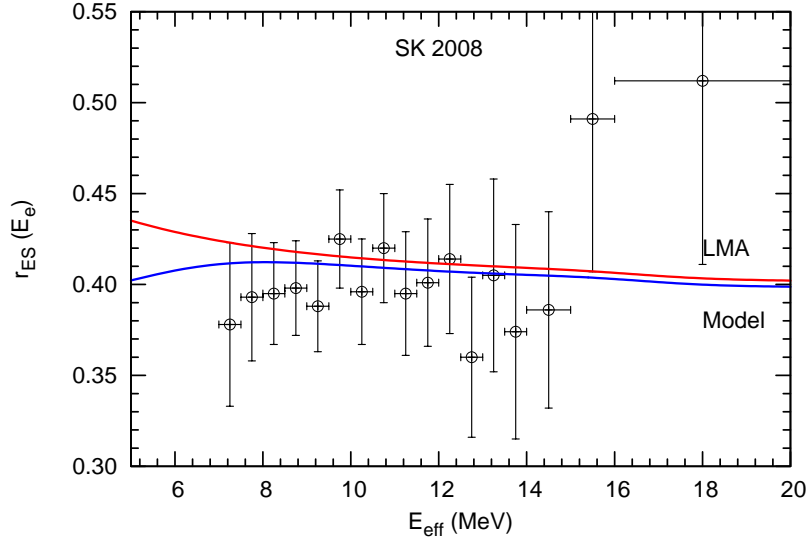}
\vspace{-0.3cm}
\caption{The same as fig.3 for the 2008 data \cite{Cravens:2008aa}.}
\end{figure}
Turning now to the spectral fits, we have for the SNO CC spectrum (see fig.2)
\begin{equation}
\chi^2_{{\rm CC~spectrum}}(\rm{LMA})=24.0/13~d.o.f.~~~,~~~\chi^2_{{\rm CC~spectrum}}({\rm model})=21.6/13~d.o.f
\end{equation}
where we took into account all 15 data points. As expected, the sterile neutrino model fits the data
better than LMA, as it reflects the downturn of the spectrum for the lower energies. Regarding the
ES spectrum, it should be noted that not only this appears to be flat, but there is also a hint for a 
downturn at the lower energies (see figs.3 and 4). To this end we performed as well a $\chi^2$ analysis
for both SK data sets. For the 2010 data we find \footnote{In eqs.(12) and (13) we have not taken into account 
the highest energy data points in the calculation in view of their poorer statistics.}
\begin{equation}
\chi^2_{{\rm ES~spectrum}}(\rm{LMA})=19.2/16~d.o.f.~~~,~~~\chi^2_{{\rm ES~spectrum}}({\rm model})=19.5/16~d.o.f
\end{equation}
so that the two fits look similar for this set. For the 2008 data on the other hand
\begin{equation}
\chi^2_{{\rm ES~spectrum}}(\rm{LMA})=3.6/12~d.o.f.~~~,~~~\chi^2_{{\rm ES~spectrum}}({\rm model})=2.6/12~d.o.f
\end{equation}
and thus a better fit for the sterile model, as expected. Note that the smaller magnitude of the $\chi^2$'s 
in the 2008 data is a consequence of both the smaller number of degrees of freedom and the fact that the predictions, 
especially the sterile model one, lie practically all within the data error bars, which is by no means the case for 
the 2010 data set.

We next summarize our conclusions:
\begin{itemize}
\item We still need to fill the gap in our knowledge of the solar neutrino survival probability in the
intermediate energy range, the vacuum matter transition region.
\item The LMA prediction seems to point in the wrong direction at the low energy end of the 
electron spectra, especially the charged current one.
\item Oscillations to a sterile neutrino which is almost degenerate with the active ones with $\Delta m^2_{41}=10^{-5}eV^2$
and $sin\theta_{14}=0.04$ seem to provide a solution to these inconsistencies.
\item Inserting such a sterile neutrino, which would be the 5th or 6th neutrino, adds to the already confusing situation 
concerning sterile neutrino scenarios.
\item However from the experience with neutrino oscillations, we have learned that a confusing picture at the start may 
eventually emerge, after an accumulation of experimental tests for several years, as a clear and positive one. This
may well be the case with the sterile neutrino extensions of the electroweak standard model.
\end{itemize}

\section*{Acknowledgments}

 C.R.~Das acknowledges a scholarship
from the Funda\c{c}\~{a}o para a Ci\^{e}ncia e a Tecnologia (FCT,
Portugal) (ref. SFRH/BPD/41091/2007), also greatly thanks the Department of Physics,
Jyv\"askyl\"{a} University, in particular Prof. Jukka Maalampi (HOD) for
hospitality and financial support.
This work was partially
supported by FCT through the projects CERN/FP/123580/2011, 
PTDC/FIS-NUC/0548/2012 and CFTP-FCT Unit 777 (PEst-OE/FIS/UI0777/2013)
which are partially funded through POCTI (FEDER).

\end{document}